\documentclass{article}
\usepackage{spconf,amsmath,graphicx,amssymb}
\usepackage{subcaption}
\usepackage[ruled,linesnumbered]{algorithm2e}
\usepackage{soul}

\usepackage{multirow}

\usepackage[activate]{microtype}
\usepackage[dvipsnames]{xcolor}

\DeclareMathOperator{\EX}{\mathbb{E}}

\title{Pre-training in Deep Reinforcement Learning for \\Automatic Speech Recognition}
\name{Thejan Rajapakshe$^{1}$, Rajib Rana$^{1}$, Siddique Latif$^{1,2}$, Sara Khalifa$^{2}$, Bj\"{o}rn W.\ Schuller$^{3,4}$}

\address{ $^1$University of Southern Queensland, Australia\\
  $^2$Distributed Sensing Systems Group, Data61, CSIRO Australia\\
  $^3$GLAM -- Group on Language, Audio \& Music, Imperial College London, UK\\
  $^4$ZD.B Chair of Embedded Intelligence for Health Care \& Wellbeing, University of Augsburg, Germany}
\begin{document}
%
\maketitle
\begin{abstract}
Deep reinforcement learning (deep RL) is a combination of deep learning with reinforcement learning principles to create efficient methods that can learn by interacting with its environment. This led to breakthroughs in many complex tasks that were previously difficult to solve. However, deep RL requires a large amount of training time that makes it difficult to use in various real-life applications like human-computer interaction (HCI). Therefore, in this paper, we study 
pre-training in deep RL to reduce the training time and improve the performance in speech recognition, a popular application of HCI. We achieve significantly improved performance in less time on 
a publicly available speech command recognition dataset.

\end{abstract}
\section{Introduction}
Reinforcement
Learning (RL) follows the principle of behaviourist psychology and learns 
in a similar way as a child learns to perform a new task. RL has been 
repeatedly successful in the past~\cite{singh2002optimizing,tesauro1995temporal}, however, the successes were mostly limited to low-dimensional problems. In recent years, deep learning has significantly advanced the field of RL, with the use of deep learning algorithms within RL giving rise to the field of ``deep reinforcement learning''. Deep learning enables RL to operate in high-dimensional state and action spaces and 
can now be used for complex decision-making problems~\cite{arulkumaran2017brief}.

Deep RL algorithms have mostly been applied to video or image processing domains that include playing video games~\cite{silver2016mastering,mnih2015human} to indoor navigation~\cite{zhu2017target}. Only a very limited number of studies have explored the promising aspects of deep RL in the field of audio processing in particular for speech processing. 
In this paper, we study this under-researched topic. In particular, we conduct a case study of the feasibility of deep RL for 
automatic 
speech command classification. 

A major challenge of deep RL is that it often requires a prohibitively large amount of training time and data to reach reasonable performance, making it inapplicable in real-world settings~\cite{cruz2017pre}. Leveraging humans to provide demonstrations (known as learning from demonstration (LfD) in RL has recently gained  traction as a possible way of speeding up deep RL~\cite{vinyals2017starcraft,hester2018deep,kurin2017atari}. In  LfD, 
actions demonstrated by the human are considered as the ground truth labels for a given input game/image frame. 
An agent closely simulates the demonstrator's policy at the start, and later on 
learns to surpass the demonstrator~\cite{cruz2017pre}. However, LfD holds a distinct challenge, in the sense that it often requires the agent to acquire skills from only a few demonstrations and interactions due to the time and expense of acquiring them~\cite{Calinon2018}. Therefore,
LfDs are generally not scalable for especially high-dimensional problems. 

Pre-training the underlying deep neural network (in Section~\ref{sec:methods} we discuss the structure of RL in detail) is another approach to speed up training in deep RL. In~\cite{abtahi2011deep}, the
authors combine Deep Belief Networks (DBNs) with RL to take benifit of the unsupervised pre-training phase in DBNs, and then use the DBN as the opening point for a neural network function approximator. Furthermore, in~\cite{anderson2015faster}, the
authors demonstrate 
that a pre-trained hidden layer architecture can reduce the time required to solve reinforcement learning problems. While these studies show the promise of using pre-trained deep neural 
networks in non-audio domains, the feasibility of pre-training is not well understood for 
the audio field in general. 

We found very few studies in audio using RL/deep RL. In~\cite{stockholm2009reinforcement}, the 
authors describe 
an avenue of using RL to classify audio files into several mood classes depending upon listener response during a performance. In~\cite{lakomkin2018emorl}, 
the authors introduce the `EmoRL' model that triggers an emotion classifier as soon as it gains enough confidence while listening to an emotional speech. The authors cast this problem into a RL problem by training an emotion classification agent to perform two actions: wait and terminate. The agent selects the terminate action to stop processing incoming speech and classify it based on the observation. The objective was to achieve a trade-off between performance (accuracy) and latency by punishing wrong classifications actions as well as too delayed predictions through the reward function. While 
the authors in the above studies use RL for audio, they do not have a focus on pre-training to improve the performance of deep RL.

In this paper, 
we propose pre-training for improving 
the performance and speed of Deep RL while conducting speech classification. Results from the case study with the 
recent public Speech Commands Dataset \cite{warden_speech_2018} show that pre-training offers significant improvement in accuracy and helps achieve faster convergence.

\section{Methodology}
\label{sec:methods}
In this study, we investigate the feasibility of pre-training in RL algorithms for speech recognition. We present the details of the proposed model in this section.

\subsection{Speech Command Recognition Model}
\begin{figure}[t!]
    \centering
    \includegraphics[width=\linewidth]{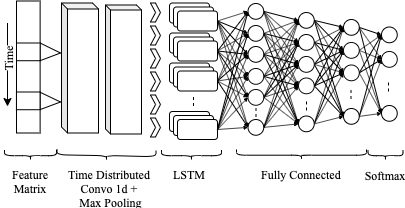}
    \caption{Speech command recognition model architecture.}
    \label{fig:speech_command_model}
\end{figure}

The considered policy network model consists of a speech command recognition model as shown in  figure~\ref{fig:speech_command_model}.
Considering the fact that Convolutional Neural Networks (CNNs) and Long Short Term Memory (LSTM) Recurrent Neural Networks (RNNs) can be combined to improve the performance \cite{latif2019direct}, 
we assembled CNN layers on top of an LSTM RNN layer. LSTM RNNs are good at learning the temporal structure of a feature map \cite{latif2018phonocardiographic}, and CNNs are strong in diminishing frequency variations \cite{rana2019multi}. The outputs from the LSTM RNN layer are passed on to fully connected layers to learn discriminative features during training \cite{latif2019direct}. In this way, our proposed policy network is empowered by convolutional layers for learning high-level abstraction, an LSTM layer to capture long-term temporal context, and finally fully connected layers for learning discriminative representation. 



\subsection{Deep Reinforcement Learning Framework}

\begin{figure}[h]
    \centering
    \includegraphics[width=\linewidth]{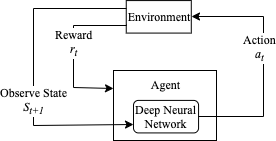}
    \caption{Framework of the proposed Deep RL. }
    \label{fig:rl_framework}
\end{figure}

The reinforcement learning framework mainly consists of two major entities namely ``agent'' and ``environment''. The \textit{action} decided by the agent is executed on the environment and it notifies the agent with the reward and next state in the environment. In this work, we focus on deep RL that involves a Deep Neural Network (DNN) structure in the agent module to resolve the action taken by observing the state which is illustrated in Figure \ref{fig:rl_framework}. We modelled this problem as a Markov decision process (MDP) \cite{tetreault_reinforcement_2008}. This can be considered as a tuple $(S,A,P,R)$, where $S$ is the state space, $A$ is the action space, $P$ is the state transition policy, and $R$ is the reward function. Since the core goal of this problem is classification, we modelled the MDP in such a way that the predicted classes are to be as actions, $A$, and the states, $S$ are the features of each audio segment in a batch of size $\eta$. An action decision is carried out by an RL agent which receives a reward ($r_{t}$) using the following reward function: 
\begin{equation}
    r_{t}=
    \begin{cases}
      +1, & \text{if}\ a_{t} = g_{t} \\
      -1, & \text{otherwise}, 
    \end{cases}
\end{equation}

where $g_{t}$ is the ground truth value of the specific speech utterance. We modelled the probability of actions using the following equation: 
\begin{equation}
    a_{t} = \text{Softmax}(W^{a}\cdot S_{t} +b^{a}),
\end{equation}
where $a_{t}$ is the action selection probability, and $W^{a}$ and $b^{a}$ are the weight and bias values. $S_{t}$ is the output from the previous hidden layer. The softmax function is  defined as:

\begin{equation}
    \text{Softmax}(\alpha_{j}) = \frac{e^{\alpha_{j}}}{\sum_{i=1}^{n} e^{\alpha_{i}} }.
\end{equation}

The target of the RL agent is to maximise the expected return in the policy
\begin{equation}
    J_{a} (\theta_a,\theta_s) = \EX_{\pi(a_{t}|s_{t}; \theta_a,\theta_s)}[r_{t}], 
\end{equation}
where $\pi(a_{t}|s_{t}; \theta_a,\theta_s)$ is the policy of agent, and $r_{t}$ is the expected reward return at state $t$.

\subsubsection{REINFORCE Training}
The REINFORCE algorithm is used to approximate the gradient to maximise the objective function $J(\theta_{a},\theta_{s})$. 

\begin{algorithm}[t!]
\SetAlgoLined
 initialise state space\;
 initialise policy network model\;
 pre-train policy network\; 
 retrieve initial state $s_{1}$\;
 \For{$i \leftarrow 1$ \KwTo $N_{E}$}{
    initialise $E_{s}, E_{a}, E_{r}$\;
    \While{!d}{
        $a_{i} \leftarrow $ get action($s_{i}$)\;
        $s_{i+1}$, $r_{i}$, $d \leftarrow execute(a_{i}$)\;
        $E_{s} \leftarrow s_{i} + E_{s}$\;
        $E_{r} \leftarrow r_{i} + E_{r}$\;
        $E_{a} \leftarrow a_{i} + E_{a}$\;
    }
    $train(E_{s}, E_{a}, E_{r})$\;
 }
 \caption{REINFORCE algorithm implementation}
 \label{alg:rl_algo}
\end{algorithm}

Algorithm~\ref{alg:rl_algo} describes the algorithmic steps followed throughout the RL action prediction process, where $N_{E}$ indicates the maximum number of episodes to run (10,000 experiments), $S_{i}$ is the state at instant $i$, $a_{i}$ is the predicted action for the $s_{i}$ at $i^{th}$ instant, $r_{i}$ is the reward obtained by executing the predicted action $a_{i}$, $d$ is a boolean flag indicating the end of an episode, where the end of the episode is decided when $i$ reaches the step size $\eta$ (50). $E_{s}, E_{a}, E_{r}$ are arrays collecting the values of $s_{i}$, $a_{i}$, $r_{i}$ for each step, which is consumed by the policy model's training method $train$. 

For a given set of examples in the state space $S$, initially, the environment sends the $s_{1}$ to the RL agent. The RL agent infers the corresponding action probabilities through the policy network and selects the highest probable action $a_{1}$ and returns it to the environment. The environment then calculates the reward $r_{1}$ for the action-state combination and returns to the agent with the reward and next state $s_{t+1}$. Each $r_{i}$, $a_{i}$, and $s_{i}$ are stored, and then, the policy network is trained with those values in an episode.


\section{Experimental Setup}

\subsection{Datasets}


The Speech Commands Dataset \cite{warden_speech_2018} is an audio corpus of 105,829 utterances containing 30 command keywords spoken by 2,618 speakers. Each utterance of a one-second file is stored in the `.wav' file format with 16\,kHz sampling rate. This dataset contains mainly two subsets of command keywords, namely ``Main Commands'', and ``Sub Commands''. Table~\ref{tab:sc_dataset} shows the distribution of the 30 keywords among the two subsets. 

\begin{table}[h]
    \centering
    \caption{Distribution of keywords in the Speech Commands Dataset}
    \begin{tabular}{p{8em}p{13em}}
        \hline
        Subset & Commands \\
        \hline
         Main Commands & one, two, three, four, five, six, seven, eight, nine, down, go, left, no, off, on, right, stop, up, yes, zero  \\\hline
         Sub Commands & bed, bird, cat, dog, happy, house, Marvin, Sheila, tree, wow
         \\\hline
    \end{tabular}
    
    \label{tab:sc_dataset}
\end{table}

\subsection{Feature Extraction}
In this study, we use Mel Frequency Cepstral Coefficients (MFCC) to represent the speech signal. MFCCs are very popular features and widely used in speech and audio analysis/processing \cite{latif2019direct,davis1980comparison}. 
We extract 40 MFCCs from the mel-spectrograms with a frame length of 2,048 and a hop length of 512 using Librosa \cite{mcfee_librosa:_2015}. 

\subsection{Model Recipe}


We use the Tensorflow library to implement the model, a combination of CNN and LSTM: The initial layers are 1d convolution layers wrapped in time distributed wrappers with filter sizes of 16 and 8,  respectively, followed by a max-pooling layer. The feature maps are then passed to an LSTM layer of 50 cells for learning the temporal features. A dropout layer of dropout rate 0.3 is used for regularisation. Finally, three fully connected layers of 512, 256, and 64 units respectively are added before the softmax layer. 
The input to the model is a matrix of $n \times f$, where $n$ is the number of MFCCs (40), and $f$ is the number of frames (87) in the MFCC spectrum. 
We use a Stochastic gradient descent optimiser 
with a learning rate of $10^{-4}$. 

The score value $V_{j}$ is defined as the sum of the rewards $r_{i}$ produced within the $j^{th}$ episode. The score variable can be utilised to infer the overall accuracy ($H$) of the RL Agent within a given episode.
\begin{equation}
    H_{i} = \frac{ V_{i} - \eta \times \text{min}(r)}{\eta \times (\text{max}(r)-\text{min}(r))} \times 100 \%.
\end{equation}

\section{Results}

Experiments were carried out focusing on the effect that pre-training has on the accuracy of the RL Agent. Three subsets of Speech Command datasets were selected, namely ``binary'', ``20 class'', and ``30 class''.  A binary subset contains only the speech commands of the ``left'' and ``right'' classes. The 20 classes and 30 classes subsets contain "Main" commands and the merge of ``Main'' and ``Sub'' commands, respectively. Each subset was experimented as without pre-training and with pre-training. 



\begin{figure*}[t!]
    \centering
    \includegraphics[trim=0.2cm 0.2cm 0.2cm 0.2cm,clip=true,width=\textwidth]{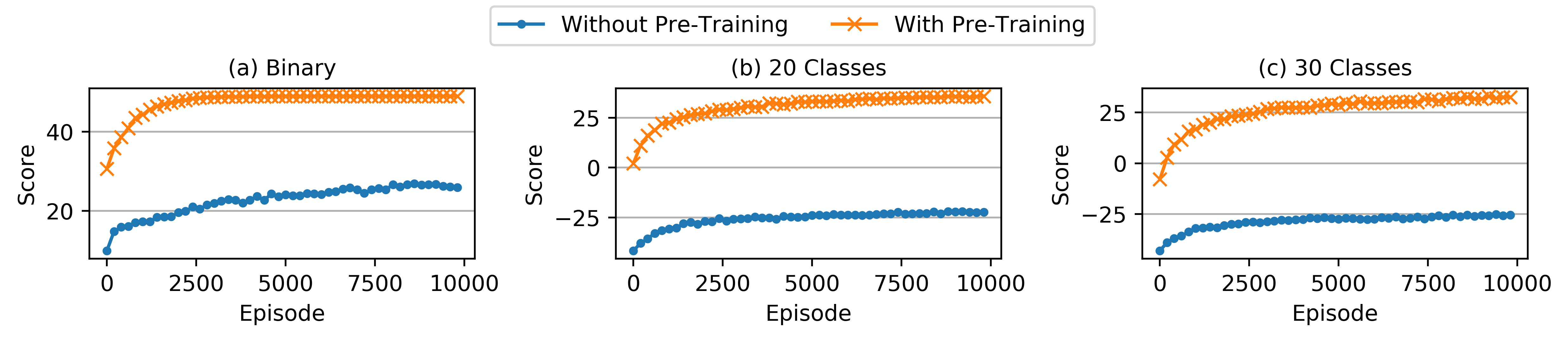}
    \caption{Performance evaluations of the model on three different scenarios}
    \label{fig:performence}
\end{figure*}

\begin{figure*}[t!]
    \centering
    \includegraphics[trim=0.2cm 0.2cm 0.2cm 0.2cm,clip=true,width=\textwidth]{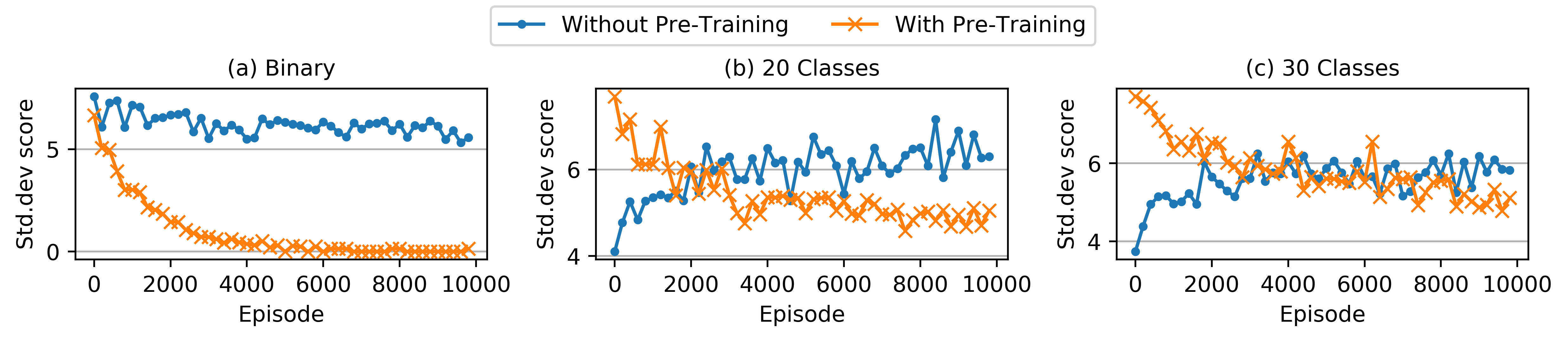}
    \caption{Standard Deviation of the score with episode on three different scenarios}
    \label{fig:std_dev_episode}
\end{figure*}



The mean score of a batch of 200 episodes is plotted against the episode number in  Figure~\ref{fig:performence}. Interpreting the graphs in Figure~\ref{fig:performence}, one can find that the pre-training increases the overall score. It can be seen that the binary class classification nearly reaches its maximum score at 50 within initial 2,500 episodes. The other 2 subsets show a score over 25 within 10,000 episodes. 


The rates of change of score (velocity of score) for the initial 500 and 1,000 episodes were calculated by equation~\ref{eq:velocity} and tabulated in Table~\ref{tab:score_velocity}.
\begin{equation}
    \text{Velocity}_x = \frac{\text{mean}(V_x:V_{x+5}) - \text{mean}(V_0:V_5)}{x}.
    \label{eq:velocity}
\end{equation}
The results in Table~\ref{tab:score_velocity} convey that the velocity of score increases by pre-training within the initial 1,000 episodes for all 3 subsets of experiments. This lets one conclude that the pre-training can decrease the time taken for the RL agent to converge to better accuracy.

\begin{table}[t!]
\centering
\scriptsize
\caption{Velocity of score change in the initial episodes.}
\label{tab:score_velocity}
\resizebox{\linewidth}{!}{%
\begin{tabular}{|l|r|r|}
\hline
\multirow{2}{*}{\# Classes} & \multicolumn{2}{c|}{Change of Velocity of score (\%)} \\ \cline{2-3} 
 & Initial 500 episodes & Initial 1000 episodes \\ \hline
2 & -0.9 & \textbf{4.4} \\ \hline
20 & \textbf{15.2} & 8.8 \\ \hline
30 & \textbf{6.4} & 11.2 \\ \hline
\end{tabular}%
}
\end{table}



\begin{table}[]
\scriptsize
\centering
\caption{Improvement of the score with pre-training.}
\resizebox{\linewidth}{!}{%
\begin{tabular}{|l|r|r|r|}
\hline
\multirow{2}{*}{\# Classes} & \multicolumn{2}{c|}{Score after 10000 episodes} & \multirow{2}{*}{Improvement (\%)} \\ \cline{2-3}
 & w/o pre-train & w/ pre-train &  \\ \hline
2 & 29.4 & 49.0 & 19.6 \\ \hline
20 & -23.8 & 37.0 & 60.8 \\ \hline
30 & -23.4 & 29.0 & 52.4 \\ \hline
\end{tabular}%
}

\label{tab:score_improvement}
\end{table}

Table~\ref{tab:score_improvement} shows the mean score of the latest 5 episodes for the ``with'' (w/) and ``without'' (w/o) pre-train scenarios. The improvement column shows the increment of score of the ``with pre-train'' with respect to ``without pre-train'' scenario, where the improvement is calculated by  equation~\ref{eq:improvement}. Each improvement is a positive improvement. This concludes that the overall final score (accuracy) of the RL agent policy network model can be improved by pre-training.

\begin{equation}
    \text{Improvement} = \frac{x_{\text{with}} - x_{\text{without}}}{\eta \times (\text{max}(r)-\text{min}(r))} \times 100 \%.
    \label{eq:improvement}
\end{equation}

According to Figure~\ref{fig:std_dev_episode}, the standard deviation of the score decreases rapidly with time in the pre-trained scenario. This again shows that  predictions of the RL agent are increasing with time and pre-training is accelerating the process.

\section{Conclusions}

In this paper, we propose the use of pre-training in deep reinforcement learning (deep RL) for speech recognition. The proposed model uses pre-training knowledge to achieve a better score while reducing the convergence time. We evaluated the proposed RL model using the speech command dataset for three different classifications scenarios, which include binary (two different speech commands), and 10 and 30 class tasks. Results show that pre-training helps to achieve considerably better results in a lower number of episodes. In future efforts, we want to study the feasibility of using unrelated data to pre-train deep RL to further improve its performance and convergence.



\end{document}